\documentclass[prb,twocolumn,superscriptaddress,byrevtex]{revtex4}
\usepackage{amsfonts}
\usepackage{amsthm}
\usepackage{graphics}

\newtheorem{theorem}{Theorem}

\begin{document}

\title{Nonequivalent ensembles and metastability}\thanks{Contribution
to the Proceedings of the 31st Workshop of the International School of Solid
State Physics ``Complexity, Metastability and Nonextensivity'', held at the
Ettore Majorana Foundation and Centre for Scientific Culture,
Erice, Sicily, Italy, July 2004. Edited by
C. Tsallis, A. Rapisarda and C. Beck. To be published by World Scientific,
2005.}

\author{Hugo Touchette}
\email{htouchet@alum.mit.edu}
\address{School of Mathematical Sciences, Queen Mary, University of London,
London E1 4NS, UK}

\author{Richard S. Ellis}
\email{rsellis@math.umass.edu}
\address{Department of Mathematics and Statistics, University of Massachusetts,
Amherst, MA 01003, USA}

\begin{abstract}
This paper reviews a number of fundamental connections that exist between
nonequivalent microcanonical and canonical ensembles, the appearance
of first-order phase transitions in the canonical ensemble, and
thermodynamic metastable behavior.
\end{abstract}

\date{\today}

\maketitle

\section{Introduction}

The goal of this short paper is to trace a line of relationships that goes
from the phenomenon of nonequivalent microcanonical and canonical ensembles
to that of thermodynamic metastability. Our approach will aim for the most
part at stressing the physics of these relationships, but care will also be
taken to formulate them in a precise mathematical language. However, due to
the limitation in available space, many mathematical details will have to be
left aside, including the proofs of all the results stated here. References
containing these proofs, when they exist, will be mentioned to assist the
reader. Another more complete paper that treats these relationships with
full mathematical details is also in preparation,\cite{ellis2005} based on
the recent doctoral dissertation of one of us.\cite{touchette2003}

Far from being exhaustive, we hope that this short review can serve as a
starting point in the literature for the reader interested in knowing about
nonequivalent ensembles, as well as those interested in phase transitions
and metastable behavior in many-body systems. These proceedings are a
testament to the fact that there remain at present many unsolved problems
related to metastability, and it is our belief that what has been learned in
studies of nonequivalent ensembles could yield useful clues for solving
these problems. Perhaps the most obvious of these clues is the fact that
many of the systems described in these pages (see, e.g., the contributions
on the HMF model) exhibit negative values of the heat capacity at fixed
energies at the same time that they exhibit metastable states. The
negativity of the heat capacity at fixed energy is well known to be related
to the nonequivalence of the microcanonical and canonical ensembles. It is
also, as we will see here, a direct indication of metastable behavior.

\section{Nonequivalent ensembles}

The equivalence of the microcanonical and canonical ensembles is most
usually explained by saying that although the canonical ensemble is not a
fixed-mean-energy ensemble like the microcanonical ensemble, it must
`converge' to a fixed-mean-energy ensemble in the thermodynamic limit, and
so must become or must realize a microcanonical ensemble in that limit.\cite{ellis2004,touchette2004} 
This explanation is not far from being entirely
valid, but there is a problem with it: the canonical ensemble may not in
fact realize at equilibrium all the mean energies that can be realized in
the microcanonical ensemble.\cite{touchette2004} In other words, the range
of the equilibrium mean energy $u_\beta $ realized in the canonical ensemble
by fixing the inverse temperature $\beta $ may be only a subset of the range
of definition of the mean energy $u$ itself. If this is the case, then the
microcanonical ensemble must be richer than the canonical ensemble because
there are values of the mean energy that can assessed within the
microcanonical ensemble, but not within the canonical ensemble. The two
ensembles must therefore be nonequivalent.

\begin{figure*}[t]
\begin{center}
\resizebox{0.9\textwidth}{!}{\includegraphics{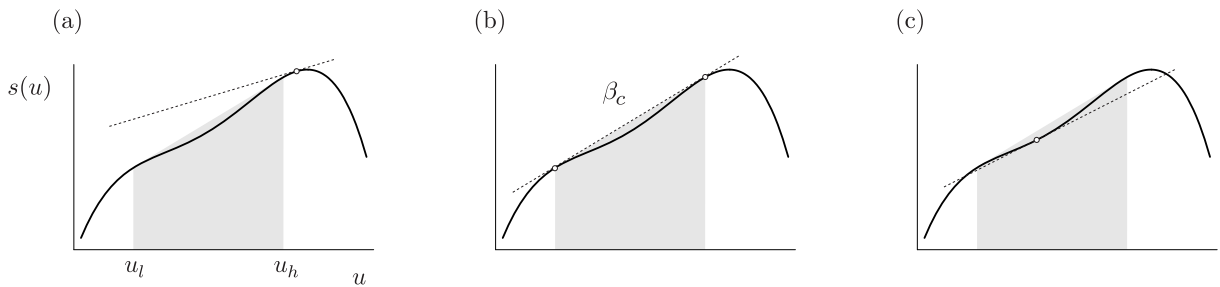}}
\end{center}
\caption{(a) Concave point of the microcanonical entropy function
$s(u)$ which admits a supporting line.
(b) Two concave points of $s(u)$ which admit the same supporting line. (c)
Nonconcave point of $s(u)$ which does not admit a supporting line.}
\label{suppline1}
\end{figure*}

To see how this possibility can arise, and how it is related in fact to the
nonconcavity of the microcanonical entropy function, let us introduce some
notation. We consider, as is usual in statistical mechanics, an $n$-body
system with Hamiltonian $U$ and mean entropy $s(u)=S(U)/n$, where $u=U/n$ is
the mean energy. To state our first result, we need to define an important
concept in convex analysis known as a \textit{supporting line}.\cite{rockafellar1970,ellis2000}
This is done as follows: we say that $s$ admits
a supporting line at $u$ if there exists $\beta \in \Bbb{R}$ such that
$s(v)\leq s(u)+\beta (v-u)$ for all admissible $u$. From a geometric point of
view, the requirement of a supporting line should be clear: it means that we
can draw a line above the graph of $s(u)$ that passes only through the point 
$(u,s(u))$; see Fig.~\ref{suppline1}. The slope of this line is $\beta $.

\begin{theorem}
\label{canequil1}Let $u_\beta $ be the value of the mean Hamiltonian
realized at equilibrium in the canonical ensemble with inverse temperature
$\beta $. (There can be more than one equilibrium value.) Then, for any
admissible mean energy value $u$, there exists $\beta $ such that $u_\beta
=u $ if and only if $s$ admits a supporting line at $u$ with slope $\beta $.
\end{theorem}

This simple result seems to have floated in the minds of physicists for a
long time. It is implicit, for example, when considering the physical
meaning of first-order phase transitions in the canonical ensemble and their
connection with nonconcave entropies.\cite{thirring1970,hertel1971,eyink1993,lynden1999,gross1997,gross2001,chomaz2001,gulminelli2002}
However, to the best of our knowledge, there has never been a clear
formulation of this result until recently.\cite{touchette2003,touchette2004}
This can be explained in part by the fact that the concept of a supporting
line is not well known in physics.

The full application of our first theorem is presented in
Fig.~\ref{suppline1}
which shows the plot of a generic entropy function having a
nonconcave part. This figure depicts three possible cases:

(a) Mean energy value $u$ for which $s$ admits a supporting line. In this
case, the value $u$ can be realized at equilibrium in the canonical ensemble
by setting $\beta $ equal to the slope of the supporting line passing
through $(u,s(u))$. We naturally expect in this case to see the
microcanonical ensemble at $u$ give the same equilibrium predictions as the
canonical ensemble at $\beta $ since the latter ensemble reduces to a
single-mean-energy ensemble with $u_\beta =u$. Note that $\beta $ must be
such that $s^{\prime }(u)=\beta $ if $s$ is differentiable.

(b) There exists a single supporting line that touches two points of the
graph of the microcanonical entropy. In this case, not one but \textit{two}
values of the mean energy---e.g., $u_l$
and $u_h$ in Fig.~\ref{suppline1}---are realized
at equilibrium in the canonical ensemble for
$\beta $ corresponding to the slope of the supporting line. This situation,
as will be clear in the next section, corresponds to a state of coexisting
phases which universally signals the onset of a first-order phase transition
in the canonical ensemble.

(c) Mean energy $u$ for which $s$ admits no supporting line. This also
applies for all $u\in (u_l,u_h)$. Theorem \ref{canequil1} states
for this case that the canonical ensemble must be
blind to the properties of the microcanonical ensemble since it cannot
realize at equilibrium any of the mean energies $u\in (u_l,u_h)$ for any
values of $\beta $. This means in particular that the standard thermodynamic
relation $\beta =s^{\prime }(u)$ ceases to be valid in this region.\cite
{ellis2004,touchette2004} The next theorem relates this case of
nonequivalent ensembles with the occurrence of negative values of the heat
capacity in the microcanonical ensemble.\cite{gross1997,gross2001}

\begin{theorem}
Define the microcanonical heat capacity at the mean energy value $u$ by
$c(u)=-s^{\prime }(u)^2s^{\prime \prime }(u)^{-1}$. If $c(u)<0$, then $s$
does not have a supporting line at $u$.
\end{theorem}

This result is a new formulation---again because the use of supporting
lines---of an old result that relates the negativity of $c$ with the
nonequivalence of the microcanonical and canonical ensembles. Usually what
is concluded is that these two ensembles must be nonequivalent when $c<0$
because the heat capacity can never be negative in the canonical ensemble.\cite{thirring1970,hertel1971,lynden1999,gross1997,gross2001}
Our
formulation has the advantage of stressing the physical root of negative
heat capacities, namely that the mean energies $u$ which are such that
$c(u)<0$ are not \textit{equilibrium} mean energies in the canonical
ensemble. This point will be discussed further in Section~\ref{secmeta}. 
For now, let us note in closing this section that the
negativity of $c$ is only a sufficient condition for ensemble
nonequivalence, not a necessary one.\cite{ellis2004,touchette2004} Thus it
is not true that the canonical ensemble is blind to the microcanonical
ensemble only for those mean energy values $u$ such that $c(u)<0$, as is
often claimed.\cite{thirring1970,hertel1971,lynden1999,gross1997,gross2001}
As we have seen, the canonical ensemble is in fact blind to the
microcanonical ensemble for all $u$ at which $s$ admits no supporting lines,
and that, in general, comprises more values $u$ than only those having
$c(u)<0$; see, e.g., Fig.~\ref{suppline1}.

\section{Nonequivalent ensembles and first-order phase transitions}

The previous section makes it clear that what is responsible for the
nonequivalence of the microcanonical and canonical ensembles is the
occurrence of a first-order phase transition in the canonical ensemble. To
be sure, just replace the word `blind' with the word `skip' to obtain a
sentence such as: the microcanonical and canonical ensembles are
nonequivalent because the canonical ensemble \textit{skips} over an interval
of mean energies which can be accessed microcanonically.\cite{thirring1970,lynden1999,gross1997,gross2001} 
The inverse temperature at
which the canonical ensemble skips over the microcanonical ensemble
corresponds, not surprisingly, to the inverse temperature at which a
first-order phase transition appears. This is the subject of the next
theorem which relates the nonconcavity property of $s(u)$ with the
differentiability property of the free energy function $\varphi (\beta )$,
the central thermodynamic quantity of the canonical ensemble which is taken
here to be defined by the limit 
\begin{equation}
\varphi (\beta )=\lim_{n\rightarrow \infty }-\frac 1n\ln Z_n(\beta ),
\end{equation}
where is $Z_n(\beta )$ is the standard $n$-body partition function.\cite{ellis2000,ellis2004}

\begin{theorem}
Assume that $s$ admits no supporting lines for all $u\in (u_l,u_h)$. Then
$\varphi $ is non-differentiable at a critical value $\beta _c$ equal to the
slope of the supporting line that bridges $u_l$ and $u_h$. The left- and
right-derivatives of $\varphi $ at $\beta _c$ equal $u_h$ and $u_l$,
respectively.
\end{theorem}

This theorem is a direct result of the fact that $\varphi (\beta )$ is the
Legendre-Fenchel transform of $s(u)$, whether $s(u)$ is concave or not, and
some basic properties of these transforms.\cite{eyink1993,ellis2000,ellis2004} 
It can be found in many works\cite{thirring1970,hertel1971,lynden1999,gross1997,gross2001} 
which do not 
define, however, the concavity of $s(u)$ in terms of supporting lines. This
is a minor omission because most of these works use an equivalent method for
defining the range of nonconcavity of $s(u)$ based on the so-called
Maxwell's construction.\cite{gross1997,gross2001} In any case, it is clear
in all the works just cited that the nonequivalence of the microcanonical
and canonical ensembles arises as a consequence of first-order phase
transitions in the canonical ensemble. The nonconcavity of $s(u)$, which
translates into a `back-bending' shape of $s^{\prime }(u)$, is in fact
sometimes taken as a definition or a probe of canonical first-order phase
transitions.\cite{gross1997,gross2001,chomaz2001,gulminelli2002} The
opposite is also possible; that is, it is possible to relate the absence of
a first-order phase transition in the canonical ensemble with the
equivalence of the microcanonical and canonical ensembles.\cite{ellis2004}
This is done in the next theorem.

\begin{theorem}
If $\varphi $ is differentiable at $\beta $, then $s$ admits a \emph{strict}
supporting line that touches the graph of $s$ only at $u=\varphi ^{\prime
}(\beta )$.
\end{theorem}

This result implies the following standard result: if $\varphi $ is
differentiable at $\beta $, then $u=\varphi ^{\prime }(\beta )$ is the
unique mean energy value realized at equilibrium in the canonical ensemble
with inverse temperature $\beta $.

\section{Nonequivalent ensembles and metastability}

\label{secmeta}

The last set of results that we will discuss directly pertains to the mean
energies which can be assessed microcanonically but not canonically. What we
want to show is that these \textit{nonequivalent} mean energies correspond
to \textit{nonequilibrium} critical mean energies of the canonical ensemble.
This is somewhat obvious given that they cannot be equilibrium mean
energies; however, what we want to discuss more specifically is the physical
nature of these nonequilibrium critical points. To do so, we have to note
that the values $u_\beta $ of the mean energy that are realized at
equilibrium in the canonical ensemble at $\beta $ are, by definition, the
global minimum of the function $F_\beta (u)=\beta u-s(u)$,\cite{chomaz2001,gulminelli2002} 
which we call the nonequilibrium free energy
function.\cite{touchette2003,touchette2004} This implies in particular that
$u_\beta $ must satisfy $\partial _uF_\beta (u_\beta )=0$, or equivalently
$\beta =s^{\prime }(u_\beta )$, assuming that $s$ is differentiable. Note
however---and this is the crucial point here---that not all the points $u$
satisfying $\beta =s^{\prime }(u)$ may globally minimize $F_\beta (u)$; some
of these critical points may actually correspond to local minimum of 
$F_\beta (u)$ or even local maximum of $F_\beta (u)$. To determine the
precise nature of these \textit{nonequilibrium canonical critical points},
we can look at the sign of the second $u$-derivative of $F_\beta (u)$ to
obtain the following result.

\begin{theorem}
Suppose that $s$ does not admit a supporting line at $u$.

\emph{(a)} If $c(u)>0$, then $u$ is a \emph{metastable} mean energy of the
canonical ensemble, in the sense that it is a local but not global minimum
of $F_\beta (u)$ for $\beta =s^{\prime }(u)$.

\emph{(b)} If $c(u)<0$, then $u$ is an \emph{unstable} mean energy of the
canonical ensemble, in the sense that it is a local maximum of $F_\beta (u)$
for $\beta =s^{\prime }(u)$.
\end{theorem}

While this result applies to the mean energy, it is interesting to see if
anything can be said about general macrostates: e.g., the magnetization or
the distribution of states. We all know, for instance, that phase
transitions in spin systems can be revealed at the level of the mean energy
(thermodynamic level) or at the level of the magnetization (macrostate
level) since both levels are related in a one-to-one fashion. Is the same
true for metastability? That is, can the metastable behavior of a system be
revealed at the macrostate level? If so, can this macrostate level of
metastability be related to the thermodynamic level of metastability defined
with respect to the mean energy?

The answer to these questions is yes, so long as we are concerned with
mean-field systems, which are basically systems for which the Hamiltonian $U$
can be expressed as a function of some macrostate $m$ of interest.\cite{touchette2003,campa2004} 
In this case, we can formulate the following
result about the macrostate values $m^u$ which are realized at equilibrium
in the microcanonical ensemble with mean energy $u$, but not in the
canonical ensemble for any $\beta $. The result is formulated in terms of
the nonequilibrium free energy $F_\beta (m)$ which is the macrostate
generalization of $F_\beta (u)$.\cite{touchette2003}

\begin{theorem}
\label{mmeta1}Suppose that $s$ does not admit a supporting line at $u$.

\emph{(a)} If $c(u)>0$, then $m^u$ is a \emph{metastable} macrostate of the
canonical ensemble, in the sense that it is a local but not global minimum
of $F_\beta (m)$ for $\beta =s^{\prime }(u)$.

\emph{(b)} If $c(u)<0$, then $m^u$ is an \emph{unstable} macrostate of the
canonical ensemble, in the sense that it is a saddle-point of $F_\beta (m)$
for $\beta =s^{\prime }(u)$.
\end{theorem}

What this results says physically is that a macrostate value $m^u$ which is
stable in the microcanonical ensemble can become unstable and thus decay in
time if we release the energy constraint and fix the inverse temperature
instead, as in the canonical ensemble. The precise way in which $m^u$ decays
in the canonical ensemble to a different equilibrium value $m_\beta $ is
determined by the local geometry of $F_\beta (m)$ around $m^u$ which is
determined, in turn, by the sign of $c(u)$. For more details on this result,
the reader is referred to two papers\cite{antoni2002,ellis2004} which
contain the result of Theorem~\ref{mmeta1} 
in a more or less conjectured form. A proof of this theorem can
be found in a recent proceedings paper of Campa and Giansanti.\cite{campa2004}
Another proof will be presented elsewhere.\cite{ellis2005}

\section{Concluding remarks}

The present paper hardly exhausts the subject of nonequivalent ensembles and
metastability. In going further, we could have reviewed recent works on the
dynamics of nonequivalent states in the canonical ensemble,\cite{antoni2004,lynden1999,bouchet2004}
as well as the dynamical stability of
these states\cite{ellis2002} which is discussed, for example, in Anteneodo's
contribution to these proceedings using an approach based on Vlasov's
equation. We could have alluded also to the fact that nonconcave entropies
are seen in fields as disconnected as string theory\cite{cobas2004} and
multifractal analysis.\cite{beck1993} Finally, we could have mentioned our
recent work on a generalization of the canonical ensemble which aims at
converting unstable and metastable states of the canonical ensemble into
stable, equilibrium states of a modified canonical ensemble so as to recover
equivalence with the microcanonical ensemble.\cite{costeniuc2004} Research
is ongoing on this topic.

\section*{Acknowledgments}

One of us (H.T.) would like to thank the organizing committee of the
Complexity, Metastability and Nonextensivity Conference for its hospitality
and for financial support. The research of H.T. was supported by NSERC
(Canada) and the Royal Society of London, while that of R.S.E. was supported
by the National Science Foundation (NSF-DMS-0202309).

\onecolumngrid

\end{document}